\documentclass[preprint,unnumsec,webpdf,modern,large,namedate]{oup-authoring-template}% uncomment this line for author year citations and comment the above
\theoremstyle{thmstyleone}%
\theoremstyle{thmstyletwo}%
\theoremstyle{thmstylethree}%

\usepackage{dsfont}

\newcommand{\pkg}[1]{{\tt #1}}

\begin{document}

\journaltitle{PREPRINT}
\DOI{PENDING}
\copyrightyear{2024}
\pubyear{2024}
\access{PREPRINT}
\appnotes{PAPER}

\firstpage{1}

%\subtitle{Subject Section}

\title[Symmetry in Boolean automata]{CANA v1.0.0 and schematodes: efficient quantification of symmetry in Boolean automata}

\author[1,2,$\dagger$]{Austin M. Marcus\ORCID{0000-0002-1611-8236}}
\author[2,$\dagger$]{Jordan Rozum\ORCID{0000-0002-4356-9809}}
\author[3]{Herbert Sizek\ORCID{0000-0003-0199-6965}}
\author[2,4,$\ast$]{Luis M. Rocha\ORCID{0000-0001-9402-887X}}

\authormark{Marcus et al.}

\address[1]{\orgdiv{Center for Complex Biological Systems}, \orgname{University of California Irvine}, \orgaddress{Irvine CA, 92697, USA}}
\address[2]{\orgdiv{Department of Systems Science and Industrial Engineering}, \orgname{Binghamton University (State University of New York)}, \orgaddress{Vestal NY, 13850, USA}}
\address[3]{\orgdiv{Department of Informatics}, \orgname{Indiana University}, \orgaddress{Bloomington IN, 47405, USA}}
\address[4]{\orgname{Instituto Gulbenkian de Ciência}, \orgaddress{Rua da Quinta Grande, 6, 2780-156 Oeiras, Portugal}}

\corresp[$\dagger$]{Equal co-author}
\corresp[$\ast$]{Corresponding author. \href{rocha@binghamton.edu}{rocha@binghamton.edu}}

% \received{Date}{0}{Year}
% \revised{Date}{0}{Year}
% \accepted{Date}{0}{Year}

%\editor{Associate Editor: Name}

%\abstract{
%\textbf{Motivation:} .\\
%\textbf{Results:} .\\
%\textbf{Availability:} .\\
%\textbf{Contact:} \href{name@email.com}{name@email.com}\\
%\textbf{Supplementary information:} Supplementary data are available at \textit{Journal Name}
%online.}

\abstract{%
% 1) want to study symmetry in biomolecular networks
The biomolecular networks underpinning cell function exhibit canalization, or the buffering of fluctuations required to function in a noisy environment. One understudied putative mechanism for canalization is the functional equivalence of a biomolecular entity's regulators (e.g., among the transcription factors for a gene). 
% 2) will use BNs to do it
We use Boolean networks to study cell regulatory systems.
%
%Cell regulatory systems are frequently studied using Boolean networks.
%
In these discrete dynamical systems, activation and inhibition of biomolecular entities (e.g., transcription of genes) are modeled as the activity of coupled 2-state automata, 
and thus the equivalence of regulators can be studied using the theory of symmetry in discrete functions. 
% 3) we have new algorithm to do it + are releasing cana 1.0.0
To this end, we present a new exact algorithm for finding maximal symmetry groups among the inputs to discrete functions. We implement this algorithm in Rust as a Python package, \pkg{schematodes}. We include \pkg{schematodes} in the new \pkg{CANA} v1.0.0 release, an open source Python library for analyzing canalization in Boolean networks, which we also present here.
% 4) we have benchmarks + case study
We compare our exact method implemented in \pkg{schematodes} to the previously published inexact method used in earlier releases of \pkg{CANA} and find that \pkg{schematodes} significantly outperforms the prior method both in speed and accuracy. We also apply \pkg{CANA} v1.0.0 to study the symmetry properties of regulatory function from an ensemble of experimentally-supported Boolean networks from the Cell Collective. 
% 5) we find substantial deviation from basline
Using \pkg{CANA} v1.0.0, we find that the distribution of a previously reported symmetry parameter, $k_s/k$, is statistically significantly different in the Cell Collective than in random automata with the same in-degree and activation bias (Kolmogorov-Smirnov test, $p<0.001$). In particular, its spread is much wider than in our null model (IQR 0.31 vs IQR 0.20 with equal medians), demonstrating that the Cell Collective is enriched in functions with extreme symmetry or asymmetry. 
%motivating future work to explore the phenotypic role of highly symmetric regulatory mechanisms.
}


% \boxedtext{
% \begin{itemize}
% \item Key boxed text here.
% \item Key boxed text here.
% \item Key boxed text here.
% \end{itemize}}

\maketitle

\section{Introduction}

% It is important to study two types of redundancy in Boolean models of biological processes. Both can be thought of as a type of symmetry. The first is an invariance in the output of the function to bit flips, and the second is an invariance in the output when variables are permuted. CANA was implemented some time ago to study these forms of redundancy. Since then, CANA has been used in a number of studies to gain insights into ``real" biological networks as well as the criticality hypothesis.

% People use BNs, we'll talk about 2 pkgs to study them: 1 new, 1 old but majorly upgraded. We'll focus on symmetry upgrades.
Boolean networks are discrete dynamical models used in social, physical, and biological sciences, and are especially popular in systems biology where threshold behaviors are common and data required to fit detailed mechanistic models are rare. Here, we discuss two software packages for their analysis. The first is an updated software package, \pkg{CANA} version 1.0.0 (v1.0.0), a Python library for Boolean network analysis with an emphasis on understanding symmetry and redundancy. The second, \pkg{schematodes}, is a new open source Python library written in Rust that underlies the symmetry computations provided in \pkg{CANA} v1.0.0. \pkg{CANA} v1.0.0 is a substantial upgrade to \pkg{CANA} version 0.1.2 (v0.1.2) \citep{correia_cana_2018,gates_effective_2021}. In addition to large speed improvements, \pkg{CANA} v1.0.0 includes new functionality for the computation of perturbation response, interaction graphs, and symmetry properties (see Supporting Materials for a brief overview of features). We optimized several \pkg{CANA} v0.1.2 functions using \pkg{Cython} and Rust with \pkg{PyO3} bindings, and in some cases, have redesigned the underlying algorithms using novel approaches with improved complexity, scaling, and accuracy. The largest improvements, and our emphasis here, involve the symmetry computations that describe the effect of permuting a Boolean function's non-redundant inputs.

% Canalization is important, we can study it better now.
In biomolecular regulatory systems, canalization (the buffering of genetic, epigenetic, and environmental fluctuations) plays a key role in establishing a robust mapping from genotype to phenotype \citep{waddington1942canalization}. Canalization requires dynamical redundancy, which manifests in several ways, including: i) multiple pathways along which a signal propagates, ii) multiple combinations of transcription factors that bind a gene's promoter region, and iii) threshold behaviors that allow depletion of one signal to be overcome by overabundance of another. To study canalization, \pkg{CANA} provides routines that quantify various types of redundancy using mathematically rigorous measures. The development of these measures began in the 2000s by extending the theory of total symmetry of Boolean functions \citep{mccluskey_detection_1956} to compress the prime implicants of a Boolean function into a set of \emph{schemata} where symmetry is described with symbols for groups of inputs that can permute \cite{marques-pita_canalization_2013}. \pkg{CANA} was first released in 2018 as v0.0.2-alpha by \cite{correia_cana_2018} and updated to v0.1.2 by \cite{gates_effective_2021}. Since then, we have made significant improvements and additions to these tools, culminating in the release of \pkg{CANA} v1.0.0.

% We give better formal grounding + algo for old methods
Here, we provide a rigorous group-theoretic description and justification for the symmetry schema redescription of \cite{marques-pita_canalization_2013}. Using this, we develop a new algorithm, \pkg{schematodes}, for prime implicant compression. The \pkg{schematodes} algorithm is implemented as a Python library written in Rust with \pkg{PyO3} bindings. Notably, the \pkg{CANA} v0.1.2 permutation symmetry calculations extrapolate from a sample subset of possible permutations and can overestimate the permutability of input signals. In contrast, the \pkg{schematodes} algorithm presented here and used in the updated \pkg{CANA} v1.0.0 is exact. Despite its exactness, the \pkg{schematodes} computation is dramatically faster overall than the previous method. 
The \pkg{schematodes} algorithm can be used to study symmetry in ensembles of biomolecular network models, as we demonstrate with 74 experimentally-supported models from the Cell Collective \citep{helikar_cell_2012} as a test-bed.
We discover that these real-world models exhibit more extreme symmetry and asymmetry than random null models, with moderately symmetric functions being underrepresented.

\section{Two-symbol schemata theory}
% \subsection{Partial symmetry and two-symbol schemata}
% quick mention of what 2-symbol symmetry is + point to SM
In this section, we briefly discuss the two-symbol schemata of \cite{marques-pita_canalization_2013} and give a more formal, comprehensive treatment in Supporting Materials. Computation of two-symbol symmetry in \pkg{CANA} identifies conditions when a function's inputs can permute without altering its output. 
A first set of symbols indicates input values and ``don't care'' inputs, while a second indicates literals in a prime implicant that permute to obtain another prime implicant.
Biologically, the amount of compression achievable (by reducing redundancy) is related to the extent of regulatory functional equivalence, which is relevant in phenomena such as genome duplication, compensatory mutation, and drug resistance \cite{marques-pita_canalization_2013,gates_effective_2021}.

% total vs partial symmetry
Some Boolean functions exhibit total symmetry, where all inputs may permute without altering the output; simple examples include the 2-input OR and AND functions (denoted by the $\lor$ and $\land$ operators respectively). As the number of inputs grows, however, such functions become relatively rare. For $k>2$ inputs, a function is more likely to be only partially symmetric, meaning a subset of its inputs permute, subject to certain constraints on their values. For example, in the function $(x_1\land x_2)\lor x_3$, the inputs do not all permute except when all inputs are equal (trivially) or exactly two inputs equal $1$. In addition, the first two inputs permute when $x_3=0$ and at least one of $x_1$ or $x_2$ is $0$. Additional trivial permutations of equal-value inputs are possible.

% partial symmetry of a set of input configurations
Formalizing these observations, consider a set of input configurations, represented as tuples (e.g., $010$ representing $x_1=0,x_2=1,x_3=0$), that all result in the same output of a regulatory function (e.g., the elements of $\{010,100,000\}$ result in $(x_1\land x_2)\lor x_3=0$). If the set is invariant under all permutations of a subset of tuple positions, the set exhibits \emph{partial symmetry} in those input positions. For example, the set $\{010,100\}\mapsto 0$ exhibits partial symmetry in the first two input positions. Sometimes, inputs within non-overlapping subsets can be independently permuted. For example, the arbitrary permutations of inputs $x_1$ and $x_2$ as well as arbitrary permutations of inputs $x_3$ and $x_4$ might independently conserve the output of the function (e.g., as in $(x_1 \land x_2) \lor \neg x_3 \lor \neg x_4$). Such cases also fall under the umbrella of partial symmetry (e.g., $\{0101,1010,0110,1001\}\mapsto 0$ exhibits partial symmetry in the first two inputs and in the last two inputs). We are particularly interested in the case when a permutations is not just a trivial reordering of inputs with the same value. That is, we seek permutation groups for which only the identify permutation maps every input configuration to itself; these are called \emph{faithful} group actions (see Supporting Materials for a formal definition). 
% maximize + decorate -> 2-symbol schema
We identify the maximal sets of input configurations that are invariant under such a partial symmetry group using a new exact algorithm implemented in Rust as a Python package \pkg{schematodes} that is integrated into \pkg{CANA} v1.0.0 (see Supporting Materials for full details of the algorithm used).
%We identify the largest \emph{faithful} sets of input configurations (tuples) that exhibit partial symmetry for a given regulatory function. 
%
%We identify the largest sets of input configurations (tuples) that exhibit non-trivial partial symmetry (a faithful group action) for a given regulatory function. 
% 

Following \cite{marques-pita_canalization_2013}, we annotate the identified maximal sets using the two-symbol schema notation in which the permuting input subsets of a representative input configuration are annotated with symbols above the input values. For example, $\mathring{1}1\mathring{0}\hat{0}\hat{1}$ indicates that the set of interest contains the 5-tuple $11001$ as well as those obtained by permuting the first and third entries and/or the fourth and fifth entries, resulting in the set $\{01101,01110,11001,11010\}$.
In other words, this set of input configurations can be compressed into the annotated schema. The ability to compress input configurations this way denotes that the underlying function possesses partial \textit{symmetry redundancy} among subsets of inputs (e.g. function will behave the same way as long as one of the last two inputs is ON and the other is OFF, no matter which).
Maximality requires that there are no larger partial symmetries that contain $\mathring{1}1\mathring{0}\hat{0}\hat{1}$ as a subset, meaning for example, that $\mathring{1}\mathring{1}\mathring{0}\hat{0}\hat{1}$ contains an input configuration that results in a different function output. Note that this procedure, illustrated here for the alphabet $\{0,1\}$, generalizes in a straightforward way to arbitrary finite alphabets. In this way, the approach of \cite{marques-pita_canalization_2013} is given a formal group-theoretic grounding.

% can run on raw inputs or PIs
This type of symmetry redundancy is distinct from, but related to, the redundancy captured by the number and size of a function's prime implicants. In its latest release, \pkg{CANA} can calculate two-symbol schemata using either the uncompressed activating input configurations or the set of prime implicants (also called one-symbol schemata in this context). In this latter case, two-symbol schemata are calculated for tuples of 0s, 1s, and \#s, where the \# symbol denotes a ``wildcard'', or unspecified input value. This reveals how much redundancy exists due to permutation symmetry within the prime implicants, separating the effects of irrelevant inputs from the effects of input symmetry~\citep{manicka_role_2017}. 
%A high degree of partial symmetry corresponds to a small number of two-symbol schemata, each representing a large subset of input configurations mapping to the same output. Functions with this property exhibit a high level of redundancy. 

% stuff to compute from 2-symbol schema
From two-symbol schemata, \pkg{CANA} can calculate various quantities and representations. For instance, the dynamic canalization map (DCM), introduced by \cite{marques-pita_canalization_2013}, represents a Boolean network as a threshold network with the necessary and sufficient control logic revealed after redundancy removal. By leveraging permutation symmetry, the DCM is generally more compact than other representations such as logic hypergraphs or parity-expanded networks \citep{wang_elementary_2011,klarner_computing_2015,rozum_parity_2021}. \pkg{CANA} can also use the two-symbol schemata to quantify the extent of partial symmetry in a function. By default, \pkg{CANA} v.1.0.0 computes the average number of nontrivially permuting inputs for each prime implicant associated with a given input configuration, and then averages over all input configurations to obtain the input symmetry, denoted $k_s$, but  alternative aggregations, such as maximum rather than average, are available.

% In \cite{} [Rion 2018], $k_s$ was defined as,
% \begin{equation}
%     k_s(x_i) = \frac{\sum\limits_{f_\alpha \in F_i} \max\limits_{\theta:f_\alpha \in \Theta_\theta} n_\theta^\circ}{|F_i|}.
% \end{equation}

% In implementation, we assumed that $k_s$ was equivalent to the sum of the input level symmetry, $s_{ji}$, defined as:
% \begin{equation}
%     s_{ji} = \frac{\sum\limits_{f_\alpha \in F_i} \underset{\theta:f_\alpha \in \Theta_\theta^i}{\avg} j\rightarrowtail \mathring{\;\;}_\theta}{|F_i|}
% \end{equation}
% Further analysis showed that $k_s(x_i) = \sum\limits_{j} s_{ji}$ is only true if the kernel operator is $\avg$. Thus, we altered CANA to calculate $k_s$ and $s_{ji}$ separately. This allows for the kernel operator (e.g. $\min$, $\max$, $\avg$) to be changed depending on the application.

\section{Symmetry in random and Cell Collective models}
\begin{figure*}
    \centering
    \includegraphics[width=0.4\linewidth]{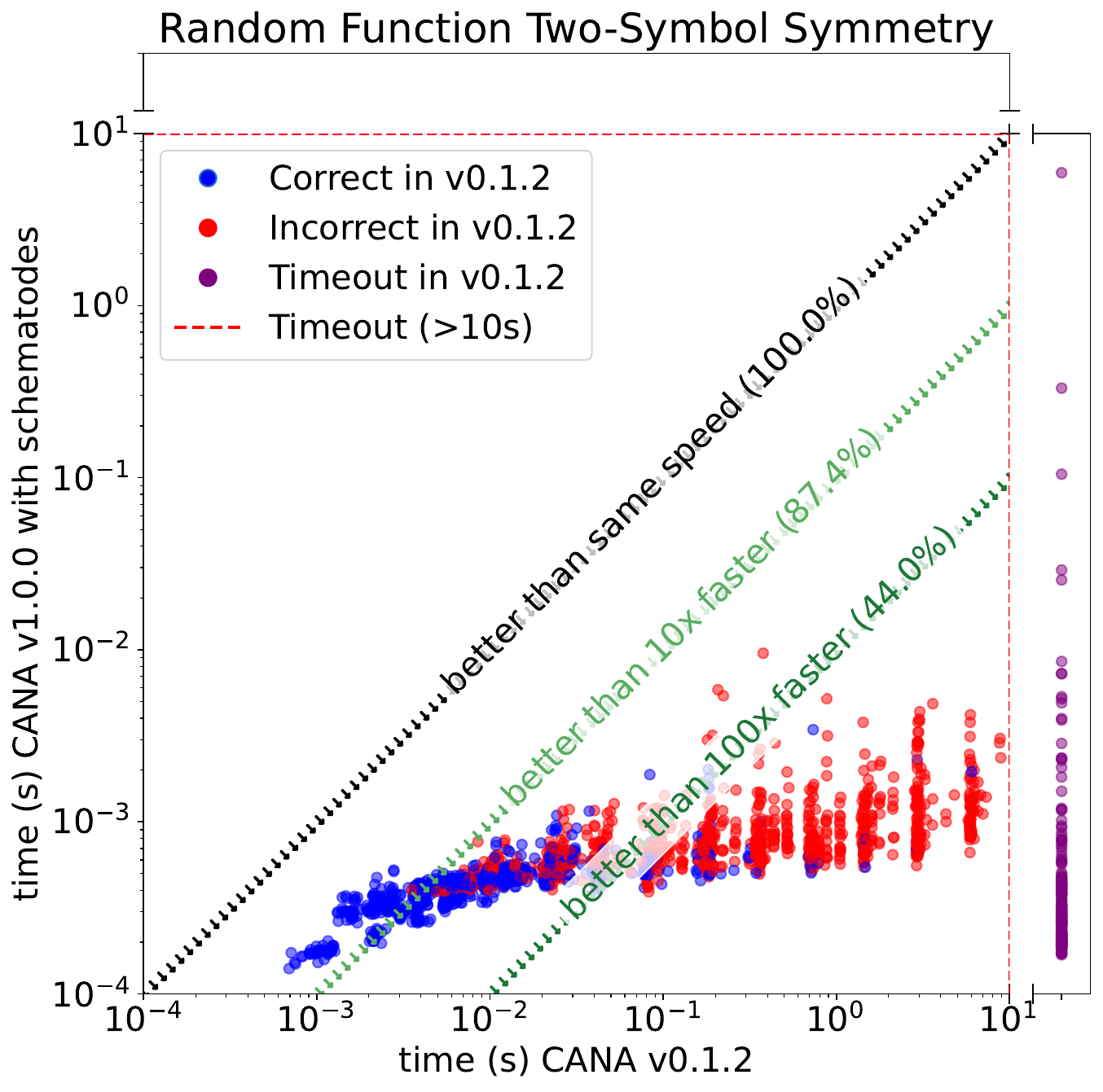}\includegraphics[width=0.6\linewidth]{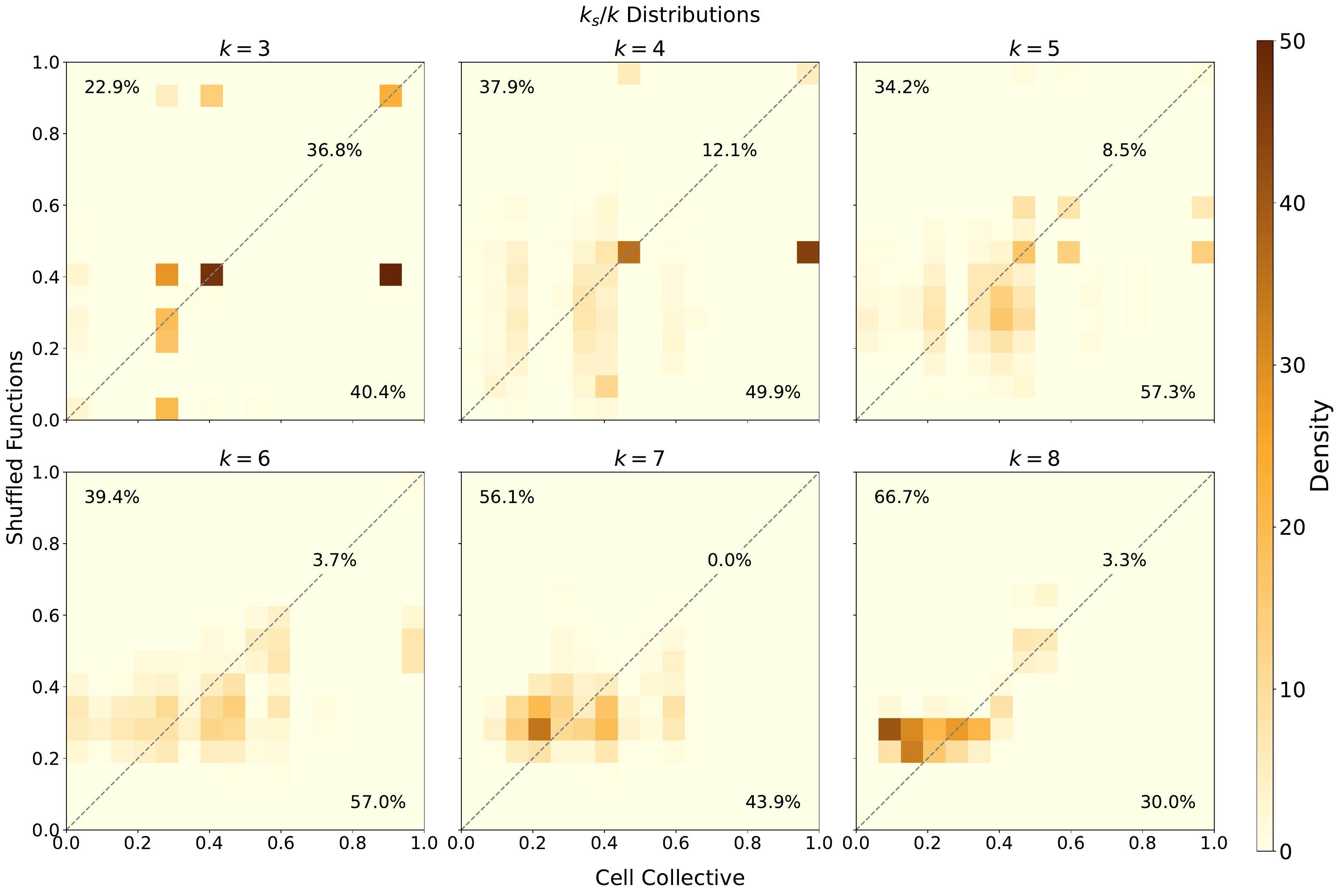}
    \caption{Benchmark timings for random automata (left) and symmetry analysis for random and Cell Collective automata (right). In the left panel, the axes show the computation time using \pkg{CANA} v1.0.0 with \pkg{schematodes} (vertical) and \pkg{CANA} v0.1.2 (horizontal). All outputs from \pkg{CANA} v1.0.0 with \pkg{schematodes} were verified to be correct. Correct (incorrect) outputs generated using the heuristic method of \pkg{CANA} v0.1.2 are shown in blue (red). Benchmarks were run on a Dell XPS 13 9380 with a 3.9GHz quad core Intel core i5 and 8GB RAM. The right panels show comparisons of symmetry measures ($k_s$) for Cell Collective functions before (horizontal axes) and after (vertical axes) shuffling, organized and normalized by in-degree ($k$). Each automaton in the Cell Collective is represented 12 times and is used to produce a random ``shuffled'' function with the same number of input configurations resulting in the ON state. Percentages in the top left, top right, and bottom right of each panel indicate how many shuffles result in increased, the same, or decreased symmetry, respectively. }
    \label{fig:main-figure}
\end{figure*}

% 1) we calculated tss & k_s on CC + RBNs 2) compared speed & accuracy 3) overview
We computed two-symbol schemata and input symmetry $k_s$ for an ensemble of randomly generated functions with five 
%(possibly redundant)
inputs to estimate symmetry redundancy in random automata (1,943 functions generated with biases, or fraction of activating input configurations, ranging from 0 to 1)---input redundancy and its dual effective connectivity has been studied in the systems biology models and random ensembles elsewhere \cite{gates_effective_2021,manicka2022effective,costa2023effective}. We also computed $k_s$ for the regulatory functions of 74 experimentally-supported Boolean networks from the Cell Collective \citep{helikar_cell_2012}.
The left panel of Figure~\ref{fig:main-figure} compares the accuracy and speed of \pkg{CANA} v1.0.0 with that of \pkg{CANA} v0.1.2 in random automata. We assessed accuracy by comparing exhaustive function evaluation to the input configurations obtained by expanding the two-symbol schemata that correspond to the $1$ (or $0$) output. The \pkg{CANA} v0.1.2 heuristic algorithm produces incorrect two-symbol schemata in 58 of the unique Cell Collective functions (12\%) and in 53\% of the randomly generated test ensemble. As an exact algorithm, the \pkg{schematodes} in \pkg{CANA} v1.0.0 achieves 100\% accuracy in completed computations (it completed all computations in the random ensemble in under 10 seconds). Moreover, it dramatically improves the calculation speed for the majority of functions (10x faster or better in 87.4\% and 100x or better in 44.0\% of random 5-input functions tested). Functions without large speed improvements generally have larger symmetry groups (more tuple elements permute). 

% highlighted difference: CC doesn't like 0<->1 permutations, only 0<-># or 1<-># permutations
Most functions in the Cell Collective are invariant under permutation of only a single $0$ or $1$ with wildcards, $\#$. Such patterns arise from nesting conjunctions or disjunctions (variables connected only by AND or OR operators, respectively), which can be viewed as threshold functions with the largest ($k$) or smallest ($1$) threshold, respectively. For example, $x_1\lor x_2\lor x_3$ is a threshold function with threshold $1$ and two-symbol schema $\mathring{1}\mathring{\#}\mathring{\#}\mapsto1$ and $000\mapsto0$. 
Notably, threshold functions with intermediate thresholds permute more than just a single $0$ or $1$ with wildcards (e.g., $x_1\land x_2\lor x_2\land x_3 \lor x_1\land x_3$ has two-symbol schemata $\mathring{1}\mathring{1}\mathring{\#}\mapsto1$ and $\mathring{0}\mathring{0}\mathring{\#}\mapsto0$). 
% Notably, the symmetry patterns we observe in the Cell Collective do not occur for threshold functions with intermediate thresholds (e.g., the 3-input function that activates if at least two of its inputs are active). 
%
Only 11 unique functions in the Cell Collective (2\% of the unique functions) have partial symmetry groups that allow a $0$ to be permuted with a $1$ (or vice versa), as opposed to with only wildcards; in contrast, two thirds of our randomly generated benchmark ensemble have this property. This is consistent with the over-representation of monotonic (unate) functions in the Cell Collective. Monotonic functions are those functions with the property that every input is unambiguously an activator or an inhibitor. Using \pkg{CANA} v1.0.0, we discovered that more than 90\% of Cell Collective models contain only monotonic functions.

% comparison to shuffled CC functions: found CC has broader distributions & bands of missing values
To assess whether the symmetry distribution observed in the Cell Collective merely reflects the expected distributions of the number of regulators and activation bias in random automata,
we shuffled the output column of the truth tables of 
%the Cell Collective update functions. 
%
%For 
each Cell Collective function to produce twelve random rules.
The resulting set of random functions have the same number of inputs, in-degree denoted by $k$ 
%(some of which may be redundant)
, and the same number of input configurations resulting in an output of $1$. 
For each of those, we computed the $k_s$ symmetry measure, normalized by $k$ and compared to the value of $k_s/k$ for the original, unshuffled function (right panels of Figure~\ref{fig:main-figure}). We note that for low in-degree ($k\leq 6$), shuffling tends to decrease symmetry (see percentages in the lower right of the right-hand panels of Figure~\ref{fig:main-figure}), whereas the opposite occurs for higher in-degree. 
%
% Because the distribution of $k_s/k$ is less sensitive to $k$ for shuffled functions, we conjecture that this may be a signature of modeler bias. 
%
Furthermore, we observe a higher spread in $k_s/k$ values for the Cell Collective functions (IQR 0.31 vs IQR 0.20 for $3\leq k\leq 8$ with median value 0.375 for both; distributions are statistically significantly different by the Kolmogorov-Smirnov test at $p<0.001$) and, for lower in-degree, an over-representation of highly symmetric functions (e.g., 75th percentile of 0.88 vs 0.42 for $k=3$). Notably, some values of $k_s$ appear much less frequently in the Cell Collective than observed in the shuffled models (vertical light bands in Fig. 1), possibly because these values are incompatible with monotonicity.

\section{Discussion}

% Further motivation for why we should study partial symmetry in biology
We have presented the open-source Python package \pkg{CANA} v1.0.0 and its novel component \pkg{schematodes} and applied them to study partial symmetry in biomolecular networks. Various theoretical mechanical and evolutionary models suggest that partial symmetry is over-represented in regulatory mechanisms. For example, \cite{grefenstette_analysis_2006} show how increasing rates of symmetry arise from the mechanical interactions between biological signaling components and the duplication and modification of binding domains. Models of genome duplication in yeast \citep{anholt_genetic_2023,conant_functional_2006} suggest that exact symmetry among duplicated elements is only lost over very long timescales, with functional partitioning. Partial symmetry may provide a fruitful avenue for studying the gradual loss of symmetry following genome duplication. Symmetry in regulatory functions may also help to explain the observed robustness of cell systems to perturbations observed by \cite{park_models_2023}.

% canalization measures are useful, ours is different (and now on good formal footing with fast code)
Measures of canalization have provided valuable insight into the interplay between redundancy and robustness in biological networks. Previous research has focused on measures derived from prime implicants \citep{marques-pita_canalization_2013, gates_effective_2021}, monotone functions \citep{grefenstette_analysis_2006}, or single node perturbations \citep{shmulevich2004activities}. \pkg{CANA} v1.0.0 and \pkg{schematodes} allow for fast and accurate computation of symmetry-based canalization measures. 

% reminder of key results in our case study
Our analysis of partial symmetry in the Cell Collective aligns with the frequently observed tendency toward monotonicity in regulatory functions and suggests a preference for nesting very stringent or lenient threshold functions over intermediate ones. At intermediate numbers of inputs ($k=5$, $k=6$, and $k=7$), the distribution of the symmetry measure $k_s$ suggests that experimentally-supported regulatory functions have a broader distribution of permutation symmetry than is expected in random models that control for input number and bias. Taken together, these results show how regulatory functions in the Cell Collective are enriched in logical extremes.

% summarize utility for future work
\pkg{CANA} v1.0.0 and \pkg{schematodes} rely on our development of a strong formal foundation for previously used symmetry measures, as an additional dimension of the canalization phenomenon in automata networks. The new  code has also enabled generalization of symmetry measurement beyond Boolean networks to multi-state automata, which we will apply in future work. The partial symmetry computations our new tools enable are essential for a full characterization of canalization in biological networks, using discrete dynamical models of regulation and signalling. Thus, \pkg{CANA} v1.0.0 and \pkg{schematodes} present an opportunity to study and  provide new explanations for the robust function of biomolecular systems. 

%USE THE BELOW OPTIONS IN CASE YOU NEED AUTHOR YEAR FORMAT.
\bibliographystyle{abbrvnat}
\bibliography{references}

\end{document}

\maketitle

\section{Overview of Other CANA Features}
\pkg{CANA} v0.1.2 was release in 2018, and its features were described in~\cite{correia_cana_2018}. In the main text, we have focused on new contributions in the area of Boolean symmetry, but \pkg{CANA} also has many tools for attractor analysis, control, redundancy computations, and network construction that were provided in the 2018 release or have been added in the intervening years. In addition to new features, we have incorporated more rigorous unit-testing and continuous integration into the \pkg{CANA} development workflow, incorporated multiple bugfixes and optimizations, and improved readability of the code and documentation. In this section, we briefly describe some of the features implemented in \pkg{CANA} that were not emphasized in the main text. 

{\it Attractor detection, simulation, and control.} \pkg{CANA} provides an interface to the BNS attractor-detection tool for finding the attractors of a Boolean network under asynchronous update~\cite{dubrova_sat-based_2011}, which uses an efficient SAT-based method. We also implement a feedback vertex set-based attractor control algorithm. \pkg{CANA} additionally provides basic brute-force algorithms for building the state transition graph, determining the fate of initial conditions, and testing the effects of node-level interventions. These brute-force methods are implemented for synchronous update. Though not implemented in \pkg{CANA} directly, we have previously presented \pkg{cubewalkers} \cite{park_models_2023}, which is interoperable with \pkg{CANA} and allows for GPU-accelerated simulation of network dynamics under various update schemes. Moreover, it allows for computation of various dynamical quantities, discussed in ~\cite{park_models_2023}.

{\it Redundancy measures and related concepts.} In addition to the symmetry measures that are the focus of this work, \pkg{CANA} provides tools for computing and analyzing prime implicants of Boolean functions. We have previously presented the concepts of effective connectivity, edge effectiveness ~\cite{marques-pita_canalization_2013, gates_effective_2021}, which can be computed in \pkg{CANA}. Effective connectivity is a measure of collective canalization that is computed as the average size of the subset of inputs necessary to determine the output of a Boolean function. In \cite{gates_effective_2021}, a method was proposed to apportion effective connectivity among the incoming edges of a Boolean node, resulting in the edge-level measure of edge effectiveness.  \pkg{CANA} also includes methods for computing the sensitivity parameter of~\cite{shmulevich2004activities}. As discussed in the main text, \pkg{CANA} computes redundancy measures using a schema redescription framework. Previously, this was accomplished using a heuristic algorithm, but this has since been replaced by \pkg{schematodes}. \pkg{CANA} has methods for visualizing these schema as tables, which have previously been described in~\cite{correia_cana_2018}.

{\it Network construction and data sets.} We have included in \pkg{CANA} a library of 74 experimentally-supported models of cell processes from the Cell Collective~\cite{helikar_cell_2012}. In addition, \pkg{CANA} includes methods for generating random Boolean networks of various kinds. \pkg{CANA} is able to construct the interaction graph between Boolean nodes as a \pkg{networkx} object. The effective graph \cite{gates_effective_2021} is an edge-weighted interaction network in which edge weights are determined by edge effectiveness. It was shown by  to capture key dynamical information related to information spread and redundancy in Boolean networks. The DCM was introduced as a way to structurally represent Boolean update functions as a threshold network \cite{marques-pita_canalization_2013}, similar to hypergraph representations used by \cite{wang_elementary_2011,klarner_computing_2015,rozum_parity_2021} and others. In contrast to these other approaches, which encode only the prime implicants, the DCM additionally encodes information about the two-symbol schemata discussed in the main text. See \cite{marques-pita_canalization_2013} for details. \pkg{CANA} includes functions for constructing and plotting the DCM of a Boolean network.

\section{Formal description of two-symbol symmetry}

In this section, we develop the formal definition of two-symbol schemata,
beginning with definitions from introductory group theory. 
Readers who are familiar with the basics of group theory and group actions 
may wish to skip ahead to Definition \ref{def:symmetric_group} (symmetric group of an index subset).

\begin{definition}[group]
A group $G$ is a nonempty set with an
associative binary operation $*:G\times G\rightarrow G$ such that
i) there is an identity element $e\in G$ satisfying $e*g=g*e=g$
for all $g\in G$, ii) every $g\in G$ has an inverse element $g^{-1}\in G$
satisfying $g*g^{-1}=g^{-1}*g=e$. 
\end{definition}

For the sake of brevity, the $*$ symbol can be omitted from group
products when its presence can be inferred from context, e.g., $g*h$
can be written $gh$. Note that, in general, $gh$ need not equal
$hg$. We will be primarily concerned with smaller groups that lie
within a larger group of transformations, and so we recall the definition
of a subgroup.

\begin{definition}[subgroup]
A subset $H\subseteq G$ is a subgroup of $G$
if $H$ forms a group under the restriction of $*$ to $H$.
\end{definition}

There are various operations that one can perform on groups and their
subsets. One operation that is particularly important for understanding
partial symmetry is the product of group subsets, which is the set
formed by taking all (ordered) group products of elements from two
or more sets.

\begin{definition}[product of group subsets]
If $S$ and $T$ are subsets
of $G$, the product $ST$ is the set $ST=\left\{ st:s\in S\text{ and }t\in T\right\} $.
\end{definition}

Note that $ST$ is not necessarily a subgroup of $G$. If, however,
$S$ and $T$ are subgroups of $G$ that commute ($ST=TS$) and have
a trivial intersection ($S\cap T=\left\{ e\right\} $), then $ST$
is a subgroup of $G$ because $*$ acts on the two terms of the product
$st$ independently, i.e., $s_{0}t_{0}s_{1}t_{1}=s_{2}t_{2}$ where
$s_{2}=s_{0}s_{1}$ and $t_{2}=t_{0}t_{1}$. This is a sufficient,
but not necessary, condition for $ST$ to be a subgroup of $G$.

We consider how a group $G$ affects elements of some set of objects
$X$ through the lens of (left) group actions.

\begin{definition}[left group action]
A left group action of a group $G$
on a set $X$ is a function $\alpha:G\times X\rightarrow X$ such
that i) $\alpha\left(e,x\right)=x$ for all $x\in X$ and ii) $\alpha\left(g,\alpha\left(h,x\right)\right)=\alpha\left(g*h,x\right)$
for all $g,h\in G$ and $x\in X$.
\end{definition}

For the sake of brevity, when $\alpha$ is clear from context, $\alpha\left(g,x\right)$
can be written as $g\cdot x$ or $gx$. All group actions discussed here are left
group actions (as opposed to right group actions); thus we simply
refer to left group actions as group actions. To characterize the
group actions that are relevant for two-symbol schemata, we recall
two properties, transitivity and faithfulness, that distinguish different
types of group actions.

\begin{definition}[transitive group action]
A group action of $G$ on $X$
is transitive if for all $x,y\in X$, there exists a $g\in G$ such
that $gx=y$.
\end{definition}

\begin{definition}[faithful group action]
A group action of $G$ on $X$
is faithful if $gx=x$ for all $x\in X$ implies $g=e$ is the identity
element of $G$.
\end{definition}

Transitivity formalizes the concept of being able to reach all elements
of a set by application of the group action, regardless of where in
the set one begins. Faithfulness formalizes the concept of a group
action specifying a nontrivial effect for all elements of the group.
We must also consider the effect a (sub)group action has on a subset
of $X$. 

\begin{definition}[restriction of a group action]
A group action $\alpha$
of $G$ on $X$ restricts to a subgroup $H$ of $G$ and a subset
$Y$ of $X$ if $\alpha\left(h,y\right)\in Y$ for all $h\in H$ and
all $y\in Y$.
\end{definition}

Two-symbol symmetry is concerned with the effects of permuting inputs
to a function. Thus, we recall the definition of a symmetric (permutation)
group on a set of indices.

\begin{definition}[symmetric group of an index subset]
\label{def:symmetric_group}
The symmetric group
of an index subset $m\subseteq\left\{ 1,\ldots,k\right\} $, denoted
$S_{m}$, is the set of permutations of $m$ with the group operation
of composition. 
\end{definition}

We use two-cycle notation to denote elements of $S_{m}$; the pair
$\left(a,b\right)$ indicates that the $a^{th}$ and $b^{th}$ elements
are interchanged under the permutation. All permutations are expressible
as compositions of two-element swaps. Any symmetric group of an index
set has a natural group action on the set $X^{k}$ (the set of $k$-tuples
with entries drawn from $X$) in which $g\in S_{m}$ acts on a tuple
$x\in X$ by permuting the entries of $x$ in the indices indicated
by $g$. That is, $g=\left(a,b\right)$ acts on $x=x_{1}\ldots x_{k}$
by swapping $x_{a}$ and $x_{b}$. The action of all other permutations
follows from composing transpositions. Note that if some symbols in
$x$ are the same, some elements of $g$ will act on $x$ by mapping
$x$ to itself. Recall that if any $x$ is held fixed by all elements
of $S_{m}$, then the natural action of $S_{m}$ on $X^{k}$ is not
faithful. The natural action of $S_{m}$ will not generally restrict
to a proper subset $Y$ of $X^{k}$, so we are motivated to consider
special subgroups of $S_{m}$ that may restrict to $Y$ while retaining
a flavor of arbitrary permutation. 

\begin{definition}[partially symmetric group] 
\label{def:partially_symmetric_group}
A partially symmetric group
is a product of symmetric groups $S_{m_{1}}\ldots S_{m_{n}}$ where
each pair of index sets $m_{i}$ and $m_{j}$ are disjoint.
\end{definition}

Note that because the index sets are disjoint, and because each $S_{m_{i}}$
is a subgroup of $S_{\left\{ 1,\ldots,k\right\} }$, every partially
symmetric group is indeed a group. Furthermore, every partially symmetric
group has a similar natural group action to that of symmetric groups
that is obtained by restricting the domain of the natural group action
on $S_{\left\{ 1,\ldots,k\right\} }$ to $S_{m_{1}}\ldots S_{m_{n}}$.
Furthermore, because the index sets are disjoint, their ordering has
no meaning, and $S_{m_{1}}S_{m_{2}}=S_{m_{2}}S_{m_{1}}$, for example.

\begin{definition}[two-symbol schema]
A two-symbol schema of a set $F\subseteq X^{k}$
is a pair $\left(Z,S_{m_{1}}\ldots S_{m_{n}}\right)$, where $Z$
is a subset of $F$ and $S_{m_{1}}\ldots S_{m_{n}}$ and a partially
symmetric group satisfying the following properties: i) the natural
group action of $S_{m_{1}}\ldots S_{m_{n}}$ restricts to a faithful
and transitive group action on $Z$, and ii) $Z$ is maximal in the
sense that if there exists $\left(Z^{\prime},S_{m_{1}^{\prime}}\ldots S_{m_{n^{\prime}}^{\prime}}\right)$
satisfying property (i), then $Z$ is not a proper subset of $Z^{\prime}$.
\end{definition}

For the sake of brevity, we express $\left(Z,S_{m_{1}}\ldots S_{m_{n}}\right)$
using a representative $z\in Z$ and indicate each index set $m_{i}$
by a distinct symbols, or decorators, above the indices $m_{i}$ contains.
For example, 
\begin{equation}
\mathring{1}0\mathring{2}\hat{2}\hat{3}
\end{equation}
is short-hand for 
\begin{equation}
\left(\left\{ 10223,10232,20123,20132\right\} ,S_{\left\{ 0,2\right\} }S_{\left\{ 3,4\right\} }\right).
\end{equation}
The ``two'' in ``two-symbol schema'' refers to the fact that
this short-hand uses two symbol sets: the set $X$, and the set of
decorator symbols.

In the context of a Boolean function, $f$, we set $k$ equal to the
number of inputs to that function. We can consider $X$ to be the
set $\left\{ 0,1\right\} $, in which case we are interested in the
two-symbol schemata of the sets $F_{0}=\left\{ x\in X^{k}:f\left(x\right)=0\right\} $
and of $F_{1}=\left\{ x\in X^{k}:f\left(x\right)=1\right\} $. More
commonly, however, we consider $X=\left\{ 0,1,\#\right\} $, where
the $\#$ symbol denotes a ``wild-card'' or unspecified variable
value, and we find the two symbol schemata of the subsets $F_{1}^{\prime}$
and $F_{0}^{\prime}$ of $X^k$, which denote the set of all prime implicants 
of $f$ and its negation, respectively.

\section{Schematodes algorithm}
The main algorithm for schematodes is provided in this section. We break the main algorithm into two pieces so as to aid interpretation.
\begin{algorithm}
\caption{Main algorithm of schematodes.}
\begin{algorithmic}[1]
\Require $tuples$, a set of tuples of equal length
\Ensure $tss$, a set of two symbol symmetry schema
    \Function{Schematodes}{$tuples$}
        \State $tss \gets \emptyset$
        \State $sig \gets$ subsets of $tuples$ grouped by signature
        \For{$oss \in sig$}
            \State $tss \gets tss \cup \Call{TwoSymbols}{oss}$
        \EndFor
        \State \textbf{return} $tss$
    \EndFunction
\end{algorithmic}
\end{algorithm}

\begin{algorithm}
\caption{Two symbol symmetry calculation for sets of tuples with equal counts of each unique entry.}
\begin{algorithmic}[1]
\Require $oss$, set of equal-length tuples with equal signature
\Ensure $sym$, a set of two symbol symmetry schema
    \Function{TwoSymbols}{$oss$}
        \State $sym \gets \emptyset$
        \State $seen \gets \emptyset$
        \State $k \gets \textbf{len}(oss)$
        \State $tc \gets \{i\in1..k: \forall t_1,t_2\in oss, t_1[n]=t_2[n]\}$
        \State $T_f \gets \{(i,j)\in S_{\left\{ 1,\ldots,k\right\} }: i,j\not\in tc\}$\Comment{potentially faithful transpositions}
        \For{$z_0 \in oss$}
            \State $candidates \gets \{(i,j)\in T_f: (i,j)\cdot z_0 \in oss\}$ 
            \State $\mathcal{C} \gets \text{power set of } candidates \text{, partially ordered in decreasing size}$
            \For{$S \in \mathcal{C}$}
                \If{$\exists S^\prime\in seen : S \subseteq S^\prime$}
                    \State continue;
                \EndIf
                \State $Z \gets \{z_0\}$
                \Repeat
                    \State $size \gets |Z|$
                    \State $Z \gets \{s\cdot z: s\in S, z\in Z\}$
                \Until{$size = |Z|$}
                \If{$Z \not\subseteq oss$}\Comment{check for closure}
                    \State continue;
                \EndIf
                \If{$\exists s\in S\not =e : \forall z\in Z, s \cdot z = z$}\Comment{check for faithfulness}
                    \State continue;
                \EndIf
                \State $seen \gets seen \cup \{S\}$
                \State $G \gets \langle S\rangle$\Comment{get group generated by $S$}
                \State $sym \gets sym \cup \{(Z,G)\}$
            \EndFor
        \EndFor
    \State \textbf{return} $sym$
    \EndFunction
\end{algorithmic}
\end{algorithm}

\newpage

\bibliographystyle{abbrv}
\bibliography{references}